# Acoustic emission source localization in thin metallic plates: a single-sensor approach based on edge reflections

Arvin Ebrahimkhanlou[a], and  Salvatore Salamone[a*]

[a] Smart Structures Research Laboratory (SSRL), Department of Civil Architectural and Environmental Engineering, University of Texas at Austin, 10100 Burnet Rd, Bldg. 177, Austin, TX 78758

* corresponding author.

Email addresses: salamone@utexas.edu (S. Salamone) arvinebr@utexas.edu (A. Ebrahimkhanlou)

**Abstract**

This paper presents a new approach for acoustic emission (AE) source localization in an isotropic plate with reflecting boundaries. The approach leverages edge reflections to identify AE sources with no blind spots, by using just a single sensor. Implementation of the proposed approach involves three main steps. First, the continuous wavelet transform (CWT) and the dispersion curves are utilized to estimate the distance between an AE source and a sensor. Then, an analytical model is proposed to predict the edge reflected waves. Finally, the correlation between the experimental and the simulated waveforms is used to estimate the AE source location. Standard pencil lead break (PLB) tests are performed on an aluminum plate to validate the algorithm. Promising results are achieved and the statistics of the estimation errors are reported.

## 1. Introduction

Plate-like structures are ubiquitous in civil, marine, and aerospace structures. Examples include bridge girders, aircraft wings and fuselages, ship hulls, etc. [1,2]. Corrosion, fatigue cracking, and impacts are some of the most common type of threats to these components. Structural health monitoring (SHM) techniques that utilize piezoelectric transducers for receiving acoustic emissions (AE) in order to localize damage in plate-like structures have received significant attention [3–14]. Conventionally, these techniques use the first arrival time of AE signals detected at multiple receiving points to locate the damage. Although this approach works relatively well for simple structures, realistic structures often have geometrical features (e.g., joints, stiffeners, rivets, and multiple layers) and boundaries that generate multiple acoustic reflections. These reflections could reduce the reliability of current source localization approaches in terms of automatic damage detection. One strategy typically used to overcome this challenge is to increase the number of sensors, which can dramatically increase the system complexity and its deployment cost. In order to solve this problem, strategies that leverage the large number of reflections present in recorded AE signal have been recently proposed. For instance, Achdjian et al. [15] formulated a statistical multi-reflection model, which uses the propagated energies in the codas (tails) of at least three AE signals to localize their source. More recently, Ernst et al. [16] proposed an approach to localize AE sources on a thin metallic plate by back propagating the edge-reflected late arrivals of the first antisymmetric Lamb wave mode (i.e. the $A_0$ mode). They used a finite element model (FEM) to back propagate the velocity signal collected from a single point laser Doppler vibrometer (LDV) and reported the required computation time for each AE localization as six hours. In addition, Ciampa and Meo [17] demonstrated the potential of using edge-reflections for single-sensor localization of AE sources. They developed a data-driven algorithm, which uses previously collected wave-field data (baseline) and correlation imaging to localize AE sources. Besides edge-reflection-based techniques, modal acoustic emission is another family of techniques that reduces the number of AE sensors to overcome the high costs associated with the sensors and data acquisition channels[18–21]. According to this technique, the multimodal characteristics of AE signals in plate-like structures can be used to localize AE sources with only two sensors [21].

Despite these notable contributions, still single-sensor source localization algorithms, even for simple metallic structures, require either extensive baseline collection or intensive computations. To overcome these challenges, this paper introduces a new source localization algorithm that leverages the echoes and reverberations of the multiple Lamb wave modes present in AE signals. Specifically, thin metallic plates are considered as the proof of concept.

The proposed algorithm consists of three key steps (see Fig. 1). First, the arrival time measurements of both fundamental Lamb wave modes (i.e., $S_0$ and $A_0$) are conducted at various frequencies to estimate the distance between the AE source and the sensor (Step I). Then, an analytical model (hereafter referred to as the multipath (MP) model) is proposed to calculate the propagation paths of the AE signals and simulate their late arrival wave packets (Step II). Finally, a correlation imaging approach is used to localize the AE source (Step III).

The rest of the paper is organized as follows: section 2 introduces the source localization algorithm and discusses its theoretical aspects. Section 3 explains the experimental setup, and section 4 goes over the achieved source localization results, their accuracy, and computational cost. Finally, section 5 presents the concluding remarks. Two appendices also accompany the paper.

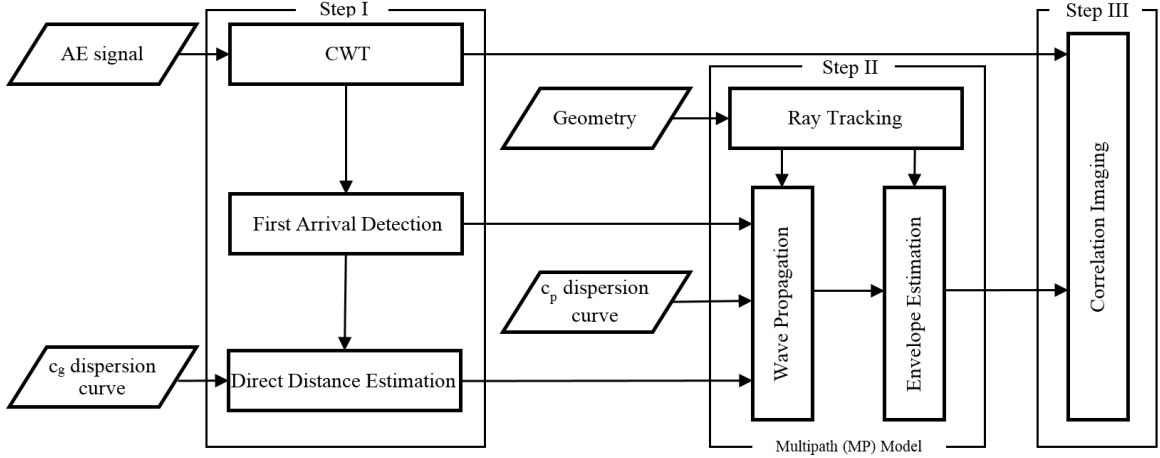

Fig. 1 Flowchart of the proposed source localization approach

## 2. Source localization algorithm

This section discusses the three steps necessary for the implementation of the proposed approach.

### 2.1. Source-to-sensor distance estimation

Consider an AE source located at distance $d$ from a sensor (see Fig. 2a). To estimate $d$, first a continuous wavelet transform (CWT) is performed on the received AE signal as:

$$C_{\text{W}}(f, t_{\text{w}}) = \frac{1}{\sqrt{s_{\text{W}}(f)}} \int_{-\infty}^{+\infty} s(t) \Psi_{\text{W}}^{*}(\frac{t - t_{\text{W}}}{s_{\text{W}}(f)}) \text{d}t \qquad (1)$$

where $s_{\text{W}}(f)$ is the non-dimensional scale parameter, $t_{\text{w}}$ is the translation parameter, and $\Psi_{\text{W}}^{*}(t)$ is the complex conjugate of the complex Morlet mother wavelet $\Psi_{\text{W}}(t)$, defined as [23]:

$$\Psi_W(t) = \frac{1}{\sqrt{\pi f_b}} \exp(2\pi f_c j t - \frac{t^2}{f_b}) \tag{2}$$

The non-dimensional parameters, $f_b$ and $f_c$, are the bandwidth parameter and central frequency, respectively. The scale parameter in Eq. (1) is defined as:

$$s_W(f) = \frac{f_c \cdot f_s}{f} \tag{3}$$

where $f_s$ is the sampling frequency of $s(t)$. The real part of the CWT can be interpreted as a Gaussian band-pass filter with its central frequency and standard deviation equal to $f$ and $f/(2\pi f_c \sqrt{f_b})$, respectively [24]; therefore the filtered signal can be represented as:

$$r(f,t) = \text{Re}(C_W(f,t_w)) \tag{4}$$

For any Lamb wave mode in $r(f,t)$ the time of flight is inversely proportional to the group velocity, $c_g(f)$. This is because the propagation distance, $d$, is the same for all frequencies and modes, that is:

$$\mathbf{c_g} \circ (\mathbf{\tau} - \tau_{AE}\mathbf{1}) = d\mathbf{1} \tag{5}$$

where $\tau_{AE}$ is the unknown origin time of the AE event, the vector $\mathbf{\tau}$ contains the arrival times of the two fundamental modes ($S_0$ and $A_0$) at different frequencies, the vector $\mathbf{c_g}$ contains their corresponding group velocities, and $\mathbf{1}$ is a vector with all elements equal to 1. The symbol ($\circ$) represents an element-wise product. To calculate the arrival times of $S_0$ and $A_0$, the Akaike information criteria (AIC) and a threshold-based approach are used, respectively (see Appendix A).

Defining $\mathbf{A} = \begin{bmatrix} \mathbf{c}_g & \mathbf{1} \end{bmatrix}$, $\mathbf{v} = \begin{bmatrix} \tau_{AE} & d_1 \end{bmatrix}^T$, and $\mathbf{b} = \mathbf{c}_g \circ \boldsymbol{\tau}$, Eq. (5) can be rearranged as a system of equations:

$$\mathbf{A}\mathbf{v} = \mathbf{b} \tag{6}$$

Since the number of equations is higher than the number of two unknowns (i.e.: $\tau_{AE}$ and $d$), the system of equations is overdetermined. The least squares (LS) method is used to solve Eq. (6):

$$\mathbf{v} = (\mathbf{A}^T\mathbf{A})^{-1}\mathbf{A}^T\mathbf{b} \tag{7}$$

When $\boldsymbol{\tau}$ is calculated for both the $S_0$ and $A_0$ modes, at least two non-parallel equations exists in $\mathbf{A}$ ( $\det(\mathbf{A}^T\mathbf{A}) \neq 0$ ). However, when only one mode is considered, the frequency $f$ should be sampled from the dispersive range of that mode; the matrix $\mathbf{A}^T\mathbf{A}$ otherwise approaches to the singularity.

## 2.2. Multipath (MP) ray tracking

Ray tracking techniques are commonly used for calculating the propagation path of waves through a medium. For instance, many studies have used them to track Lamb waves in plate-like structures [25–28]. In this paper, we propose the MP ray-tracking algorithm to trace the propagation paths of AE signals from a source to a sensor. These paths have two components known as the *direct path* and the *indirect path*. The direct path is the commonly depicted line-of-sight (i.e., the straight line) between the source and the sensor. The indirect path is the path which ends at the sensor by way of reflection from the edge of the plate. Note that there may be multiple indirect paths. In order to calculate the propagation paths, the MP algorithm needs the following parameters: (i) the dimensions of the plate, (ii) the coordinates of the sensor, (iii) an initial guess for the AE source coordinates, and (iv) the maximum number of reflections that can occur on a tracked path, $o_{\max}$. The algorithm calculates all possible paths that satisfy this maximum number. In a frequency range below the first cutoff frequency, the only propagating modes are the first symmetric ($S_0$), antisymmetric ($A_0$), and shear horizontal ($SH_0$) modes. It is assumed that at the edge of the plate: an incident $S_0$ mode reflects as $S_0$ and $SH_0$, whereas an incident $A_0$ mode reflects only as an $A_0$ without any mode conversion [29,30]. The $SH_0$ mode is not considered in this study because the AE sensors used in the experiments have negligible sensitivity to this mode. Snell's law governs the relation between the incident and reflection angles[30]:

$$k_I \sin(\theta_I) = k_R \sin(\theta_R) \tag{8}$$

where $k_I$ and $\theta_I$ are, respectively, the incident wave's wavenumber and the incident angle. Similarly, $k_R$ and $\theta_R$ are, respectively, the wavenumber and angle of the reflected wave (see Fig. 2c). Without any mode conversion, wavenumber of the incident and reflected waves are the same. Consequently, Eq. (8) requires equal incident and reflected angles. In another word, plate edges act as mirrors if no mode conversion occurs. Figure 2 visualize the overall procedure used to calculate the propagation paths from an arbitrary source to a sensor. Specifically, Fig. 2a shows the only direct path; Figure 2b shows one of the indirect paths with only a single edge reflection. To calculate this path, the source is first mirrored with respect to a reflecting edge. The figure shows the reflected source in one of the gray areas, which are the mirrored versions of the plate with respect to the four reflecting edges. Then, the line that connects the sensor and the mirrored source is considered. This line defines the propagation path until it intersects one of the edges. Finally, the intersection point is connected to the initial source location (i.e. its location before mirroring) to track the rest of the path. Fig. 2c shows the generalization of this procedure for one of the indirect paths containing two reflections from the plate's edges. Further implementation details of the MP ray-tracking algorithm can be found in an authors' previous work [27].

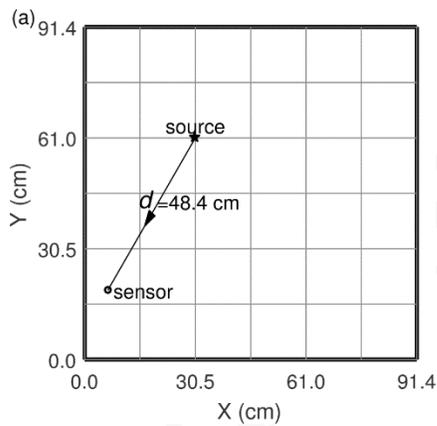

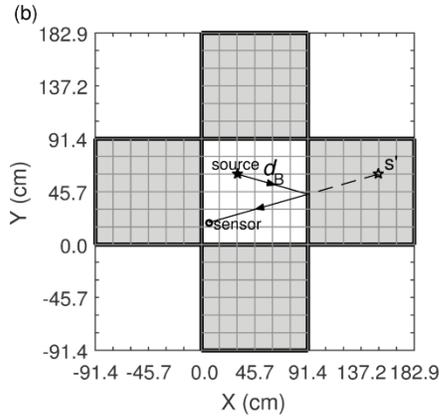

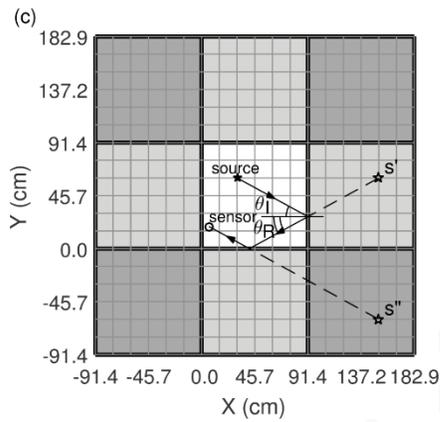

Fig. 2 The intermediate steps of the MP ray-tracking algorithm: a) a direct path, b) a path with one reflection, c) a path with two reflections

The MP ray-tracking algorithm provides all possible paths connecting an AE source to a sensor. Theoretically, there are an infinite number of such paths. Therefore, the number of reflections that can occur on each path is limited to the maximum number, $o_{max}$, specified in the input to the algorithm. Only the paths that satisfy this condition are considered. The number of such paths is defined as parameter $q$. The algorithm sorts these paths in the order of their lengths. Therefore, the first path is always the direct line connecting the source to the sensor (see Fig. 2a). For the $i^{th}$ path, the algorithm returns the length of that path, $d_i$, and the number of reflections that occur on it, $o_i$. According to this notation, $d_1 = d$, where $d$ was defined in section 2.1, and $o_1 = 0$.

## 2.3. Wave propagation

Given a function $u_0(t)$ for the out-of-plane displacement of the plate at the source, the out-of-plane displacement of an excited Lamb wave mode at a distance $d_i$ from the source can be calculated as [31]:

$$u(d_i,t) = \mathcal{F}^{-1}\{\mathcal{F}\{u_0(t)\}E(k,\omega)H_0^{(1)}(kd_i)\} \qquad (9)$$

where $E(k,\omega)$ is the excitability function of the considered mode; $H_0^{(1)}(\,\cdot\,)$ is the zero-order Hankel function of the first kind; $d_i$ is the propagation distance of the i$^{th}$ arrival; $k$ is the wavenumber of the considered mode; and $\omega$ is the angular frequency. $F\{\cdot\}$, and $F^{-1}\{\cdot\}$ are the Fourier transform and its inverse, respectively. Although Eq. (9) is valid for any Lamb wave modes, each mode needs to be considered separately. Rearranging the Eq. (9):

$$\mathcal{F}\{u_0(t)\}E(k,\omega) = \frac{\mathcal{F}\{u(d_i,t)\}}{H_0^{(1)}(kd_i)} \qquad (10)$$

Evaluating Eq. (10) at $d_1 = d$ (i.e. the direct distance from the source to the sensor) and substituting its left-hand side into Eq. (9) yields:

$$u(d_i,t) = \mathcal{F}^{-1}\{\mathcal{F}\{u(d,t)\}\frac{H_0^{(1)}(kd_i)}{H_0^{(1)}(kd)}\} \tag{11}$$

where $u(d,t)$ is the first arrival of the considered mode. For not-close-to-zero input arguments, The Hankel function can be approximated as[32]:

$$H_0^{(1)}(kd_i) = \sqrt{\frac{2}{\pi k d_i}} \exp(jkd_i - \frac{j\pi}{4}) \tag{12}$$

where j is the imaginary unit ($\sqrt{-1}$). Substituting Eq. (12) into Eq. (11) and using $k = 2\pi f / c_p(f)$:

$$u(d_i,t) = \mathcal{F}^{-1}\{(\frac{d_i}{d})^{-0.5} \mathcal{F}\{u(d,t)\}\exp(\frac{j2\pi f(d_i - d)}{c_p(f)})\} \tag{13}$$

where $c_p$ is the phase velocity dispersion curve of the considered mode.

Therefore, given $d$ from the direct distance estimation, Eq. (13) propagates a first arrival, $u(d,t)$, to a distance $d_i$ from the source. The MP ray-tracking algorithm provides the distance $d_i$ (see section 2.2). To identify the first arrival packet (i.e. $u(d,t)$), first arrival isolation methods are proposed in appendix A. These methods, which are applied to the real part of the CWT coefficient (i.e. $r(f,t)$), return the first $S_0$ and $A_0$ packets.

*2.4. Edge reflection*

The edge reflected Lamb waves can be calculated as [16]:

$$u_R(d_B, t) = (\gamma)\mathcal{F}^{-1}\{\mathcal{F}\{u_I(d_B, t)\}\exp(j\varphi)\} \quad (14)$$

where $u_I(d_B, t)$ is the incident wave; $\gamma$ is the attenuation coefficient [33]; $\varphi$ is the phase-shift; and $d_B$ is the distance from the source to the reflecting boundary (see Fig. 2b). The late arrivals of each mode are calculated by combining the Eq. (13) and Eq. (14):

$$u(d_i, t) = \gamma^{o_i}\mathcal{F}^{-1}\{(\frac{d_i}{d})^{-0.5}\mathcal{F}\{u(d,t)\}\exp(\frac{j2\pi f(d_i - d)}{c_p(f)} + j\varphi')\} \quad (15)$$

where $\varphi'$ is the overall phase-shift due to the reflections occurred on a path. Values of $d_i$ and $o_i$ are determined from the MP ray-tracking algorithm for a guessed source coordinates **x**. In this study, $\varphi'$ is assumed to be frequency independent (i.e. it shifts the wave with no distortion).

*2.5. Multipath (MP) envelope simulation model*

The MP model reconstructs the late arrival packets of filtered AE signals from their first arrivals. These filtered signals are the sum of several $S_0$ and $A_0$ wave packets that have propagated through multiple paths. Therefore, the envelope of a filtered signal can be reconstructed as:

$$\mathbf{e}(\mathbf{x}) = |\sum_{i=1}^{q}(u_{S_0}(d_i, t) + u_{A_0}(d_i, t))| \quad (16)$$

where $u_{S_0}(d_i,t)$ and $u_{A_0}(d_i,t)$ are respectively the i$^{th}$ late arrivals of the S$_0$ and A$_0$ modes that come from a source located at coordinates $\mathbf{x}$, the notation $|\cdot|$ indicates the modulus of the signal, and $q$ is the total number of paths in the MP ray-tracking algorithm (see section 2.2). To calculate $u_{S_0}(d_i,t)$ and $u_{A_0}(d_i,t)$, Eq. (15) is used with the corresponding first arrivals, $u(d_1,t)$, phase velocities, $c_p(f)$, and attenuation coefficients, $\gamma$, for the S$_0$ and A$_0$ modes. However, the phase-shift, $\varphi'$, is unknown in Eq. (15). Although $\varphi'$ is too small to affect each individual arrival packet, it can change the constructive or destructive effects of the packets on each other. To eliminate the unknown phase-shift $\varphi'$ without neglecting its effects, square root of the sum of squares (SRSS) is used instead of the modulus of summation:

$$\mathbf{e}(\mathbf{x}) = \underset{i=1}{\overset{q}{\mathrm{SRSS}}}(u_{S_0}(d_i,t), u_{A_0}(d_i,t)) \qquad (17)$$

## 2.6. Correlation imaging

Correlation imaging is a dictionary-based algorithm, which compares the similarity of experimental and simulated signals to find the most similar simulation to the experiment [27,33–35]. In a correlation image, the coordinates of pixels, $\mathbf{x}$, are initial guesses for simulating a source. According to this technique, the correlation, $\rho(\mathbf{x})$, is assigned as the value the pixel located at $\mathbf{x}$:

$$\rho(\mathbf{x}) = \frac{\sum_{i=1}^{n_s}(e_i(\mathbf{x}) - \overline{e}(\mathbf{x}))(e_i - \overline{e})}{\sqrt{\sum_{ii=1}^{n_s}(e_{ii}(\mathbf{x}) - \overline{e}(\mathbf{x}))^2 \sum_{iii=1}^{n_s}(e_{iii} - \overline{e})^2}} \qquad (18)$$

where vector $\mathbf{e}$ is the envelope of a filtered experimental signal, the vector $\mathbf{e}(\mathbf{x})$ is the envelope of the simulated signal that comes from a source at the coordinates $\mathbf{x}$ (see Eq. (17)), and $n_s$ is the length of $\mathbf{e}(\mathbf{x})$ and $\mathbf{e}$. To calculate $\mathbf{e}$, the modulus of the CWT coefficients (i.e. $|C_W(f,t_w)|$) is used, where the frequency $f$ is the same frequency used to simulate $\mathbf{e}(\mathbf{x})$. The bar on the quantities specifies their expected value. According to this technique, the pixel with the highest value is the estimated location of the source.

## 3. Experiments

To validate the proposed source localization algorithm, experiments were performed on a 91.4 cm x 91.4 cm x 0.318 cm aluminum plate. Properties of the specimen are listed in Table 1. Standard pencil lead break (PLB) tests were performed on the specimen at 64 points as shown in Fig. 3. A broadband AE sensor (Physical Acoustics PICO) located at coordinates (6.4 cm, 19.1 cm) was used to measure the AE signals. To avoid ambiguities in the localization results, the sensor was placed outside the symmetry lines of the plate. A data acquisition (DAQ) system (Mistras Micro Express) digitized the AE signals after 40dB amplification (Physical Acoustics 2/4/6 preamplifier). The low pass and high pass analog filters of the DAQ system were respectively set at 20 kHz and 400 kHz. AE signals were post-processed in MATLAB.

**Table 1.** Properties of the tested plate

| Properties | Value |
|---|---|
| Material | Aluminum alloy 6061-T6 |
| Dimension | 91.4 x 91.4 x 0.318 [cm] |
| Modulus of elasticity | 69 [GPa] |
| Poisson's ratio | 0.33 |
| Density | 2700 [Kg/m$^3$] |

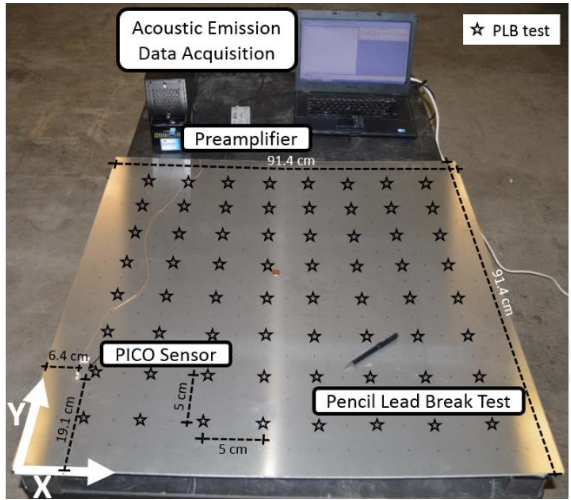

Fig. 3 Experimental setup

## 4. Results and discussion

This section presents and discusses experimental results for the intermediate steps and the overall performance of the proposed source localization algorithm. First, an AE signal generated by a PLB test is used to illustrate and validate the proposed source-to-sensor distance estimation (step I). Then, the MP simulations are discussed and compared with the same experimental signal (step II). Next, correlation imaging results are presented for three PLB tests (step III). Finally, the last two subsections use the average of 64 PLB tests to discuss the overall performance of the proposed algorithm in terms of accuracy and computation time, respectively.

*4.1. Source-to-sensor distance estimation*

Fig. 4 shows the AE signal used to validate the source-to-sensor distance estimation technique. A PLB test at the coordinates (30.5 cm, 61.0 cm) was used to generate this wideband and multimodal signal. As shown in the figure, the reference time (i.e. the time zero) of the signal was defined as the trigger time.

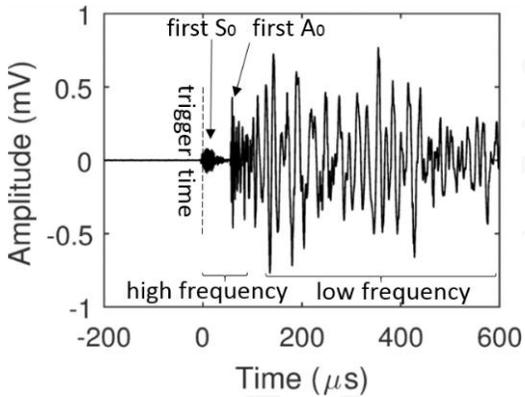

Fig. 4 AE signal generated by a PLB test at the coordinates (30.5 cm, 61.0 cm)

Fig. 5a,b visualize the CWT of the AE signal shown in Fig. 4. The non-dimensional bandwidth and central frequency parameters of the CWT were $f_b = 0.5$ and $f_c = 5$, respectively. Fig. 5a shows the modulus of the CWT coefficients for a frequency vector **f** that was uniformly sampled from 25 kHz

to 425 kHz every 1 kHz. The figure shows higher amplitudes at lower frequencies. In addition, the dispersion of the $A_0$ mode and its multiple reflections can be seen in the figure.

Fig. 5b shows the real part of the CWT coefficients at 75, 175, 275, and 375 kHz frequencies. The fundamental Lamb wave modes and several reflections can also be seen in this figure. In addition, the figure shows that the $A_0$ mode has a higher amplitude than the $S_0$ mode. The $S_0$ mode was measurable only at frequencies greater or equal to 250 kHz.

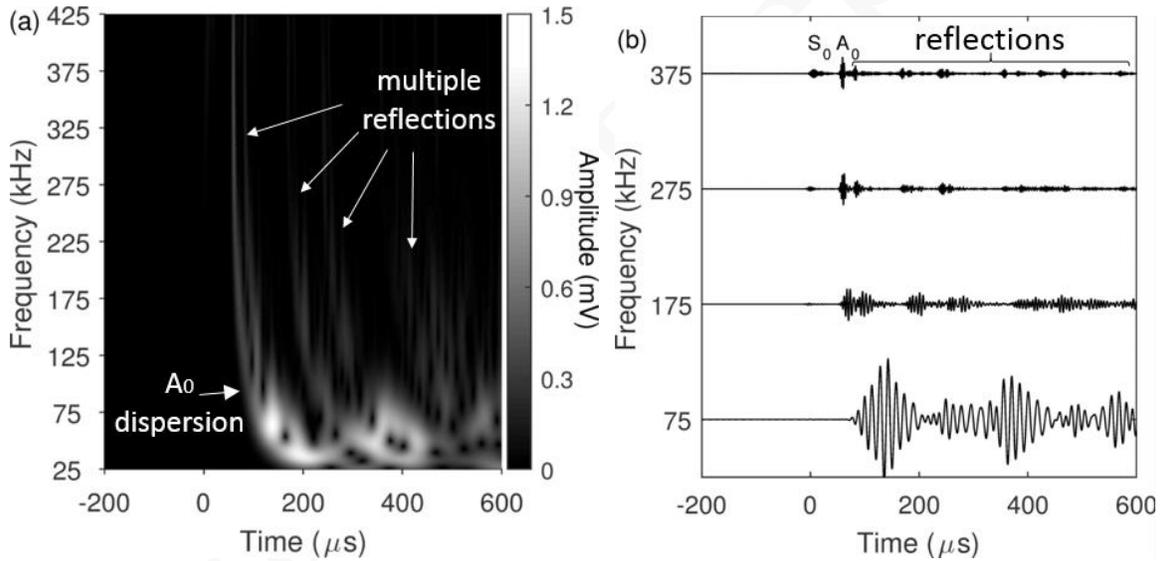

Fig. 5 The CWT of a PLB: a) the modulus of the CWT, b) the real part of the CWT at 75, 175, 275, and 375 kHz frequencies

Fig. 6 shows the first $S_0$ and $A_0$ arrival packets and their arrival time (respectively $\tau_{S_0}$ and $\tau_{A_0}$) for the real part of the same CWT at 250 kHz (i.e. $r(f,t)|_{f=250\text{ kHz}}$). The filtered signal is shown in the background, and the first arrival packets are highlighted. To isolate the first arrival packets and measure their time of arrivals, the techniques presented in Appendix A were used. In addition, the figure shows the time of AE event occurrence (i.e. $\tau_{AE}$), which was estimated by solving Eq. (5). As the figure shows, $\tau_{AE}$ is defined relative to the trigger time and thus is always a negative number.

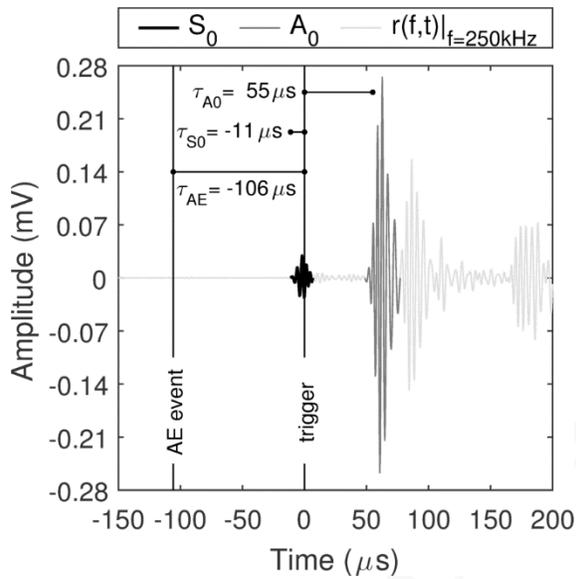

Fig. 6 A filtered AE signal at 250 kHz; the first $S_0$ and $A_0$ packets, their arrival time, and the estimated time of the AE event are shown.

Fig. 7 shows the measured $S_0$ and $A_0$ first arrival time from the real part of the same CWT. The first arrivals of the $S_0$ mode were measured at frequencies greater than 250 kHz because the $S_0$ mode had very low amplitudes at lower frequencies. For a similar reason, the first arrivals of the $A_0$ mode were measured for the frequencies less than 250 kHz. These measurements were stored in vectors $\boldsymbol{\tau}_{S_0}$ and $\boldsymbol{\tau}_{A_0}$, respectively. In addition, the corresponding group velocities of the two modes were calculated from the dispersion curves of the plate and stored in vectors $\mathbf{c}_{gS_0}$ and $\mathbf{c}_{gA_0}$, respectively. Then, the concatenations of the arrival time vectors (i.e. vector $\boldsymbol{\tau} = [\boldsymbol{\tau}_{S_0}^T, \boldsymbol{\tau}_{A_0}^T]^T$) and the group velocity vectors (i.e. $\mathbf{c}_g = [\mathbf{c}_{gS_0}^T, \mathbf{c}_{gA_0}^T]^T$) were used to construct Eq. (5). The estimated source-to-sensor distance and the occurrence time of the AE event were $d = 48.8$ cm and $\tau_{AE} = -105.7$ $\mu$s, respectively. (the actual distance was 48.4 cm). To validate the solution, the vector $\boldsymbol{\tau}$ was assumed unknown. Then, given the estimated values for $d$ and $\tau_{AE}$, Eq. (5) was solved for $\boldsymbol{\tau}$. Fig. 7 also shows these estimated values for the vector $\boldsymbol{\tau}$ and demonstrates their agreement with the measured values.

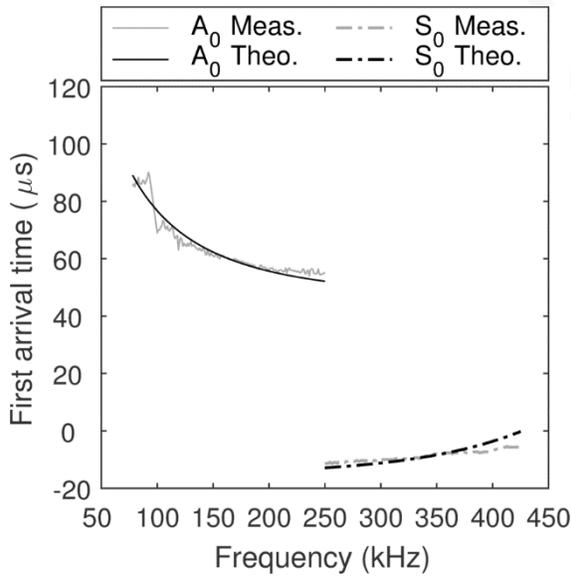

Fig. 7 Comparison of the measured and estimated values for the time of first arrivals

*4.2. Multipath (MP) model*

Fig. 8 visualizes the output of the MP ray-tracking algorithm for the source of the AE signal shown in Fig. 4. Twenty-five paths were calculated that three or fewer reflections occur on them (i.e. $q=25$ and $o_{max}=3$). For the sake of the figure's clarity, only some of the paths are shown. For each path, the travel distance, $d_i$, and the number of reflections, $o_i$, are shown. In addition, the detailed text output of the MP ray-tracking algorithm is presented in appendix B.

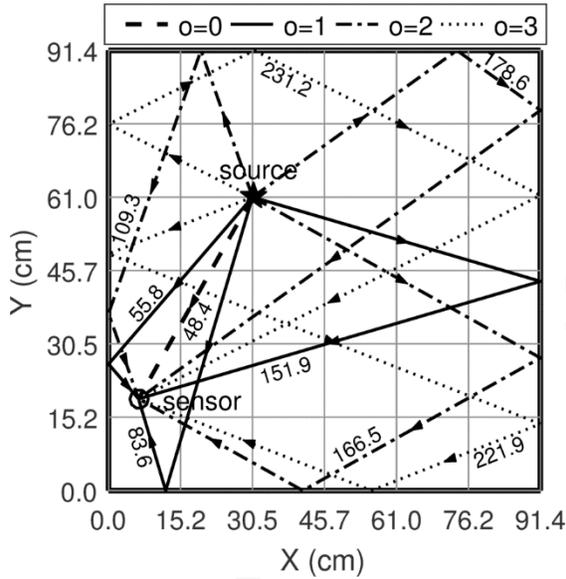

Fig. 8 The output of the MP ray-tracking algorithm for up to three reflections; the length of each path is also included in centimeters.

Fig. 9 shows the output of the wave propagation model for the isolated $A_0$ mode in Fig. 6. Time shift, attenuation, and dispersion can be seen in the figure. To simulate the propagated packets, Eq. (13)

was evaluated for 25, 50, 75, 100, and 125 cm propagation distances (i.e. $d_i$). In this equation, the estimated value for the direct source-to-sensor distance, $d$, was used (i.e. $d = 48.8$ cm).

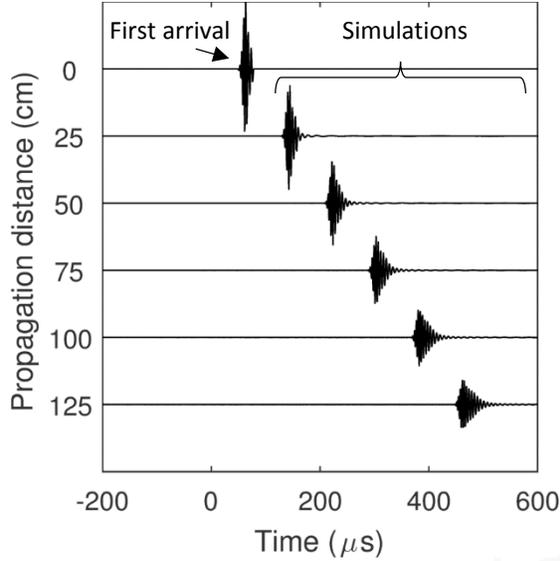

Fig. 9 Wave propagation simulations; late arrivals are reconstructed from their first arrival packets; the propagation distance is defined as $d_i - d$

Fig. 10 compares the experimental and simulated envelopes for the filtered signal shown in Fig. 6. The experimental envelope is the modulus of the CWT at 250 kHz, $\mathbf{e}$, and the simulated envelope is the output of the MP model, $\mathbf{e}(\mathbf{x})$, for the actual source location (i.e. $\mathbf{x} = [30.5 \text{ cm}, 61.0 \text{ cm}]^T$). It can be seen that the MP model can reconstruct late arrival packets from their first arrivals. In this simulation, $\gamma_{S_0} = 0.5$ and $\gamma_{A_0} = 0.8$ were used the reflections of the $S_0$ and $A_0$ modes, respectively.

Although, the simulations are not sensitive to the attenuation coefficients, higher energy loss was assumed for the $S_0$ reflections to compensate for the mode conversion of the $S_0$ mode to $SH_0$.

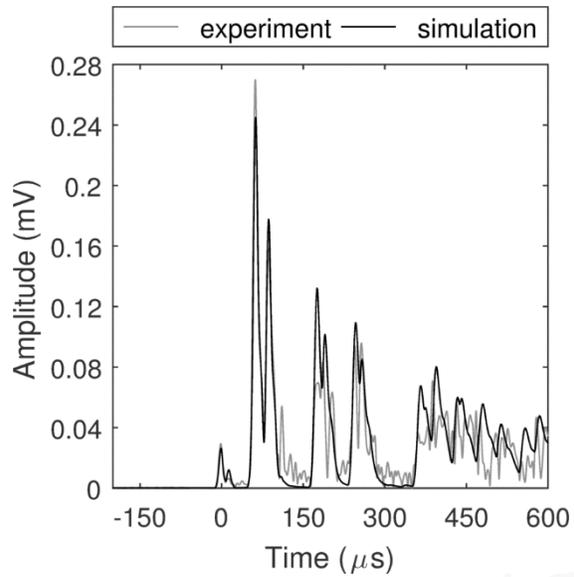

Fig. 10 Comparison between the experimental and simulated envelopes of the signal shown in Fig. 6

*4.3. Correlation imaging*

Fig. 11 shows correlation images for three PLB tests. The actual and estimated source locations can be seen in the figure. The highest correlation values are mainly located on an arc with the AE sensor at its center. In all three cases, the arc crosses the actual source and the high-value pixels have smaller variance in the radial direction (i.e. the direction of the source-to-sensor line) than the tangential direction (i.e. perpendicular to the radial direction). Fig. 11c shows a case where two maxima exist in a correlation image. Although the maxima are located closely, the one with the second highest value coincides with the actual source location.

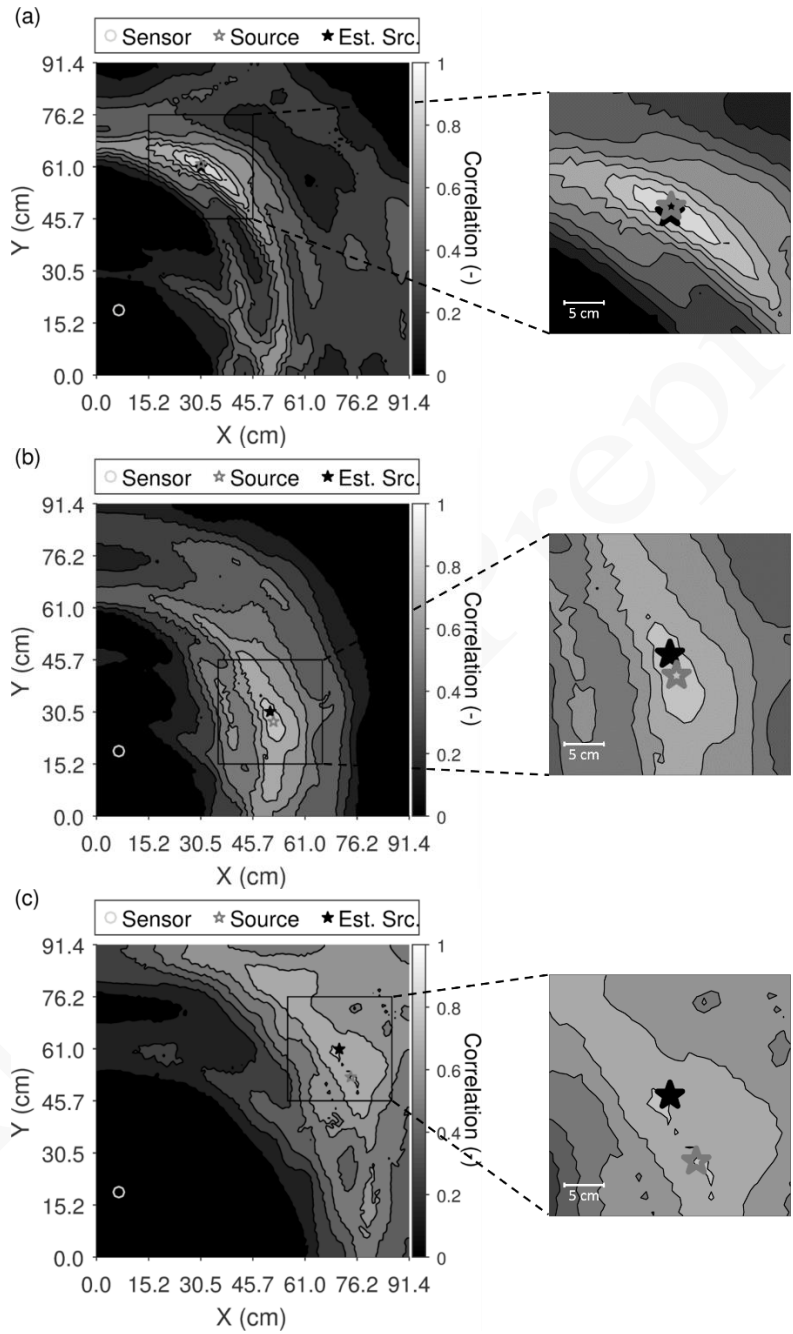
24

Fig. 11 Correlation images for PLB tests at coordinate: a) (30.5 cm, 61.0 cm), b) (50.8 cm, 30.5 cm), and c) (71.1 cm, 61.0 cm)

*4.4. Overall accuracy and error*

Fig. 12 shows the histogram of error for source-to-sensor distance estimation on the 64 PLB tests. The histogram shows less than 0.5 cm error for 26 tests. The maximum error was 2.4 cm in these estimations. In addition, the average of the absolute error was 0.9 cm, and the bias (i.e. the average error) was -0.3 cm. These results validate the source-to-sensor distance estimation step of the algorithm (step I).

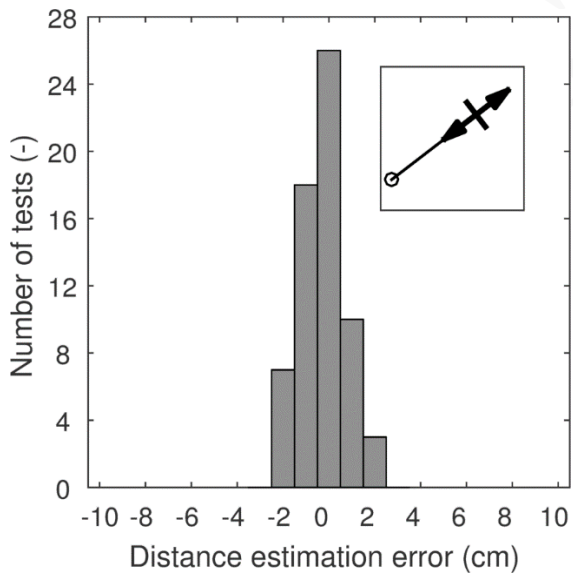

Fig. 12 Histogram of the distance estimation error for the 64 tested points

Fig. 13 compares the estimated source locations for the 64 PLB tests with their actual locations. The proposed algorithm localized all of the 64 sources. The estimated sources that had more than 5 cm error are connected to their actual source locations with a line. Overall, the maximum localization error was 8.2 cm, and the average of the absolute error was 2.9 cm. For 100x100 correlation imaging resolution, which was used to generate these results, the distance of actual sources to the closest pixel was between 0 to 1.3 cm. Therefore, at least an average of 0.6 cm error was expected. These results show that the proposed algorithm can localize AE sources without any blind zones.

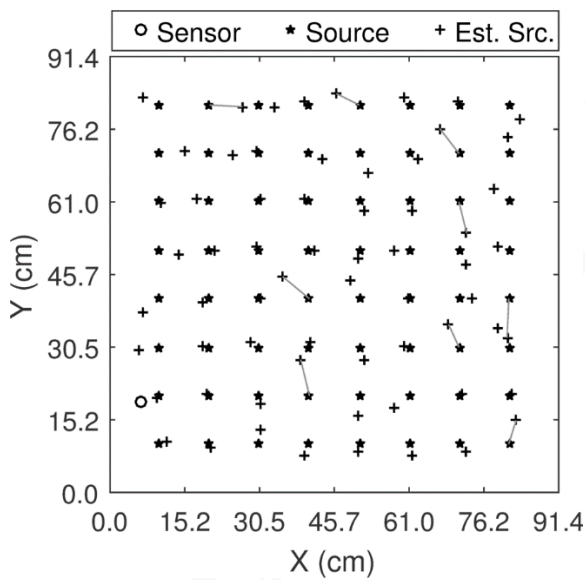

Fig. 13 Comparison of the actual and estimated source locations; for more than 5 cm error, a line connects the estimated locations to the actual ones.

Fig. 14a,b show the histogram of error in the radial and tangential directions, respectively. In the radial direction, the maximum error was 3.2 cm and the average absolute error was 0.9 cm. While, in the tangential direction, these numbers were 7.8 and 2.6 cm, respectively. Therefore, less error is expected in the radial direction than the tangential direction.

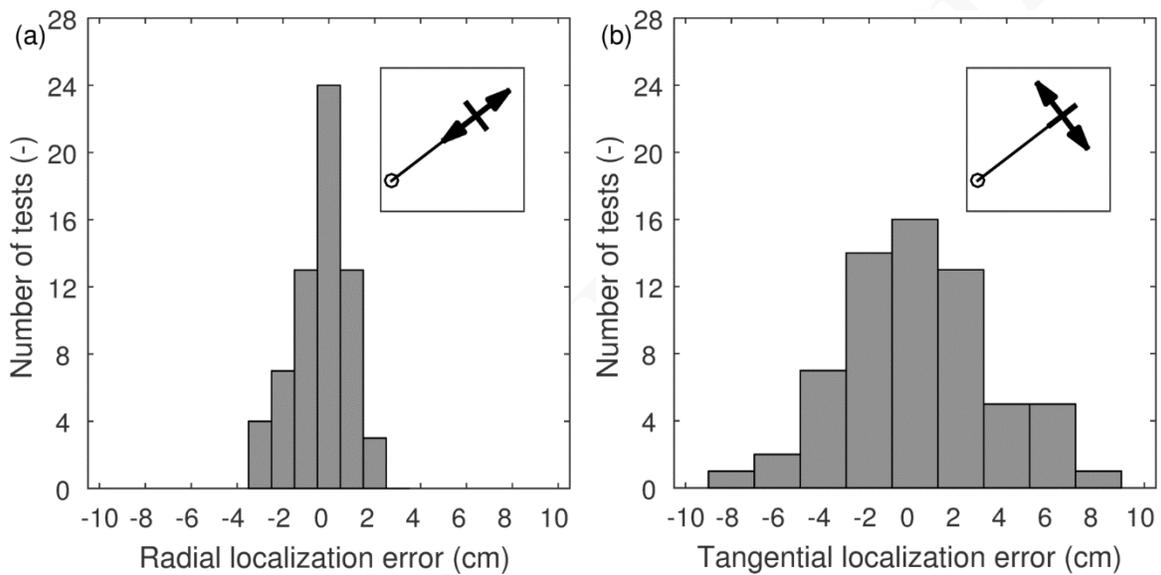

Fig. 14 Histograms of the two-dimensional localization error for the 64 tested points: a) radial direction, b) tangential direction

*4.5. Computation time*

The computation time of the proposed source localization algorithm can be broken down into the time spent on the following tasks: a) the source-to-sensor distance estimation, b) the MP ray tracking, c) the MP model and correlation imaging. A MATLAB implementation of the algorithm on a core i5

PC respectively spent 1.5 seconds, 3 minutes, and 3 seconds on average to complete the above-mentioned tasks at 100x100 pixel resolution. It needs to be noted that only one run of the MP ray-tracking algorithm is enough for the lifespan of the SHM system. From each pixel, the MP ray-tracking algorithm calculates all possible paths from that pixel to the sensor and stores them in a database. The same database can be reused for all future localizations. Therefore, the actual localization time for each AE event was 1.5+3=4.5 seconds.

## 5. Discussions and conclusions

This paper presented a novel, single-sensor AE source localization algorithm for thin metallic plates. The algorithm leverages AE reflections and reverberations as well as the multimodal nature of plate waves. Three key steps were considered. First, a least square problem was introduced to estimate the source-to-sensor distance. Then, an analytical model (the MP model) was proposed to reconstruct the edge-reflected arrivals of AE signals based on their first arrivals. Finally, the correlation analysis between the simulated and experimental signals was used to identify the AE source location. Experiments were performed on an aluminum plate to validate the approach, and very good results were achieved. It was observed that the algorithm, unlike many traditional algorithms, has no blind zones and can localize AE sources located even very close to the edges or corners of the plate. This is particularly important because those areas are potentially more prone to fatigue cracks than the rest of the plate. In addition, the accuracy and speed of the proposed approach demonstrated its potential for real-time SHM applications.

Despite the promising results presented in this paper, the proposed algorithm has some limitations. First, many plate-like structures have sophisticated geometrical features (e.g., joints, stiffeners, rivets, and multiple layers) or material properties (e.g., composite materials) that are not considered in the MP model. Second, the experiments were conducted in a laboratory setting with controlled environmental conditions. Therefore, future studies should extend the model to overcome these limitations. In addition, on-field experiments need to be conducted to verify the robustness of the approach for real applications.

## 6. Acknowledgments

This work was supported by the National Science Foundation [grant number CMMI-1333506].

**Appendix A. First arrival detection and wave packet isolation**

The subsequent subsections provide the details of the techniques used to: 1) identify the first $S_0$ and $A_0$ arrival time and 2) isolate the first arrival packets from the rest of the signal. For the $S_0$ mode, because it is the faster mode, the AIC is used [36]. Although AIC is very powerful in identifying the very first wave packet in a signal (in this case $S_0$), it is not as robust in identifying the first arrival of the $A_0$ mode. Therefore, a threshold-based technique is proposed to identify the high amplitude first arrivals of $A_0$ that come after the low amplitude arrivals of $S_0$.

*A.1. Akaike information criterion (AIC)*

The AIC is a statistical measure, which its minimum value occurs at the first arrival time of the fastest propagating mode (the $S_0$ mode in this paper) [36]:

$$AIC(t_i) = (t_i)\log(\text{var}(r_{ii})) + (t_N - t_{i+1})\log(\text{var}(r_{iii})) \tag{19}$$

where $ii \in [1, i]$, $iii \in [i+1, N]$. The parameter $N$ is the length of the signal **r**. The time at which AIC is minimized corresponds to the first arrival time Fig. A.1 shows values of AIC for the signal shown in Fig. 6.

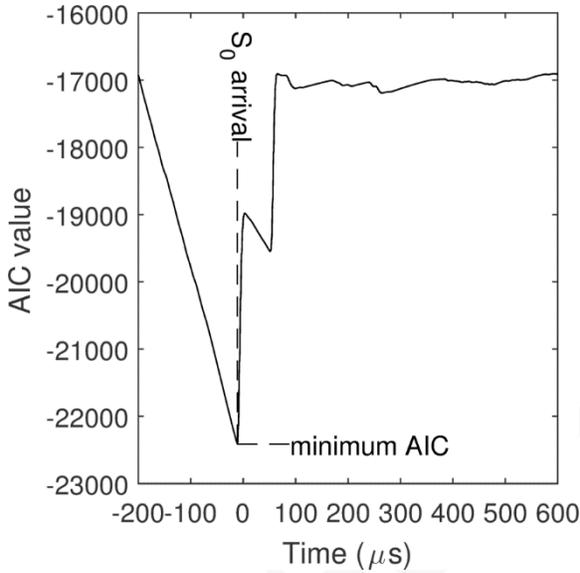

Fig. A.1 Autoregressive AIC; the minimum of AIC value corresponds to the $S_0$ arrival time

Once the first arrival time is identified, the first $S_0$ wave packet can be isolated. Fig. A.2 visualizes the isolated $S_0$ packet on the signal shown in Fig. 6. The time corresponding to the minimum of the AIC is shown as point 2. At point 4, the envelope of the signal, **e**, reaches to its first local minimum after point 2. The first $S_0$ wave packet is defined from point 1, which is the first zero crossing before point 2, to point 3, which is the last zero crossing before point 4.

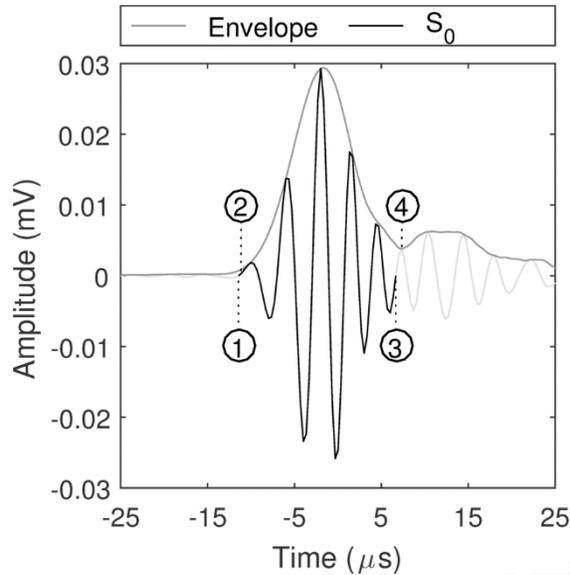

Fig. A.2 The first $S_0$ wave packet: point 1 is the first zero crossing before point 2; point 2 corresponds to the minimum of AIC; point 3 is the first zero crossing before point 4; point 4 is the local minimum of the signal's envelope.

*A.2. Threshold-base technique*

A customized threshold-base technique is used to identify the arrival time of the first $A_0$ mode. According to this technique, during the post processing, a secondary threshold is defined relative to the peak amplitude of the signal, which is calculated based on the maximum value of the signal's envelope. This secondary threshold is set at the two-third of the just defined peak amplitude. Then, a half sine is fitted to the portion of the envelope that ranges from the threshold crossing to the next adjacent peak. Finally, the zero crossing of the fitted half sine is determined by extrapolation. The time of this zero crossing defines the time of arrival. Fig. A.3 visualizes the technique on the signal shown in Fig. 6. Point 4 is the secondary threshold crossing, and point 3 is the time of arrival determined by zero crossing of the extrapolated half sine.

To isolate the first $A_0$ wave packet, the local minima of the signal's envelope are used. First, the two nearest minima before and after the time of arrival are identified (i.e., respectively, point 1 and 6 in Fig. A.3). Then, the isolated wave packet is defined from point 2, which is the first zero crossing of the signal after point 1, to point 5, which is the last zero crossing before point 6.

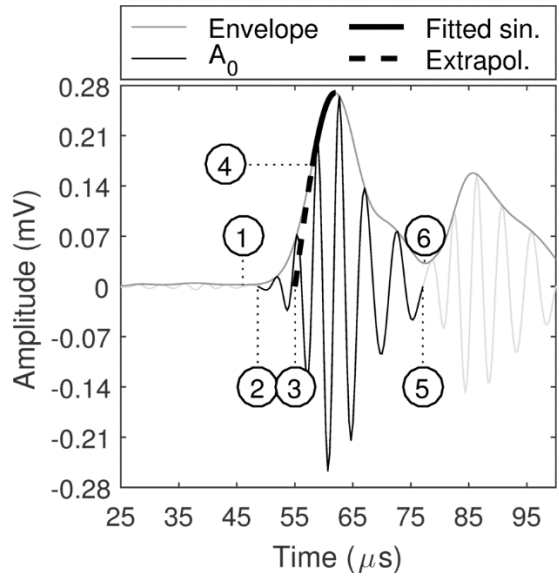

Fig. A.3 The time of arrival and the wave packet of an $A_0$ arrival: 1) the local minimum of the envelope 2) the first zero crossing after 1; 3) the considered time of arrival 4) the secondary threshold crossing; 5) the last zero crossing before 6; 6) the local minimum of the envelope

**Appendix B. Sample outputs of the MP ray-tracking algorithm**

Detailed outputs of the MP ray-tracking algorithm for the source and receiver shown in Fig. 8 are presented in Table B.1.

**Table B.1.** Output of the MP ray-tracking algorithm for the source[†] and receiver[‡] shown in Fig. 8.

| Path number ($i$) | Number of reflections ($o_i$) | Reflection sequence* | 1st reflection coordinates (cm) | 2nd reflection coordinates (cm) | 3rd reflection coordinates (cm) | Propagation distance $d_i$ (cm) |
|---|---|---|---|---|---|---|
| 1 | 0 | [] | - | - | - | 48.4 |
| 2 | 1 | [L] | ( 0.0,26.3) | - | - | 55.8 |
| 3 | 1 | [B] | (12.1, 0.0) | - | - | 83.6 |
| 4 | 2 | [L, B] | ( 0.0, 5.3) | ( 2.4, 0.0) | - | 88.1 |
| 5 | 1 | [T] | (23.3,91.4) | - | - | 105.7 |
| 6 | 2 | [L, T] | ( 0.0,36.8) | (19.6,91.4) | - | 109.3 |
| 7 | 2 | [B, T] | ( 9.6, 0.0) | (25.3,91.4) | - | 143.0 |
| 8 | 3 | [B, L, T] | ( 1.4, 0.0) | ( 0.0, 5.3) | (22.5,91.4) | 145.7 |
| 9 | 1 | [R] | (91.4,43.5) | - | - | 151.9 |
| 10 | 2 | [L, R] | ( 0.0,20.7) | (91.4,44.9) | - | 164.2 |
| 11 | 2 | [B, R] | (41.1, 0.0) | (91.4,27.6) | - | 166.5 |
| 12 | 3 | [L, B, R] | ( 0.0,15.8) | (31.4, 0.0) | (91.4,30.2) | 177.8 |
| 13 | 2 | [R, T] | (91.4,79.0) | (73.8,91.4) | - | 178.6 |
| 14 | 3 | [L, R, T] | ( 0.0,23.2) | (91.4,82.4) | (77.5,91.4) | 189.2 |
| 15 | 3 | [B, R, T] | (26.1, 0.0) | (91.4,63.1) | (62.1,91.4) | 203.0 |
| 16 | 2 | [R, L] | (91.4,36.3) | ( 0.0,54.8) | - | 211.2 |
| 17 | 3 | [B, R, L] | (55.6, 0.0) | (91.4,13.8) | ( 0.0,49.2) | 221.9 |
| 18 | 3 | [L, R, L] | ( 0.0,20.3) | (91.4,37.7) | ( 0.0,55.1) | 223.7 |
| 19 | 2 | [T, B] | (14.1,91.4) | (23.9, 0.0) | - | 226.1 |
| 20 | 3 | [L, T, B] | ( 0.0,57.8) | ( 5.5,91.4) | (20.5, 0.0) | 227.8 |
| 21 | 3 | [R, T, L] | (91.4,61.3) | (30.9,91.4) | ( 0.0,76.1) | 231.2 |
| 22 | 3 | [B, T, B] | ( 8.1, 0.0) | (16.5,91.4) | (24.9, 0.0) | 264.0 |
| 23 | 3 | [T, R, B] | (53.4,91.4) | (91.4,32.9) | (70.1, 0.0) | 268.1 |
| 24 | 3 | [T, B, T] | (12.5,91.4) | (20.2, 0.0) | (27.9,91.4) | 286.8 |
| 25 | 3 | [R, L, R] | (91.4,29.9) | ( 0.0,41.5) | (91.4,53.2) | 331.6 |

[†]Source coordinates: (30.5 cm,61.0 cm)     [‡]Sensor coordinates: (6.4 cm,19.1 cm)
* L: Left boundary   B: Bottom boundary   R: Right boundary   T: Top boundary

# 7. References

[1] T. Kundu, Acoustic source localization, Ultrasonics. 54 (2014) 25–38.

[2] L. Yu, S. Momeni, V. Godinez, V. Giurgiutiu, P. Ziehl, J. Yu, Dual Mode Sensing with Low-Profile Piezoelectric Thin Wafer Sensors for Steel Bridge Crack Detection and Diagnosis, Adv. Civ. Eng. 2012 (2012) 1–10.

[3] T. Kundu, S. Das, K. V Jata, Point of impact prediction in isotropic and anisotropic plates from the acoustic emission data, J. Acoust. Soc. Am. 122 (2007) 2057.

[4] S. Salamone, I. Bartoli, P. Di Leo, F. Lanza Di Scala, A. Ajovalasit, L. D'Acquisto, et al., High-velocity Impact Location on Aircraft Panels Using Macro-fiber Composite Piezoelectric Rosettes, J. Intell. Mater. Syst. Struct. 21 (2010) 887–896.

[5] T. Kundu, H. Nakatani, N. Takeda, Acoustic source localization in anisotropic plates, Ultrasonics. 52 (2012) 740–746.

[6] E. Dehghan Niri, S. Salamone, A probabilistic framework for acoustic emission source localization in plate-like structures, Smart Mater. Struct. 21 (2012) 35009.

[7] H. Nakatani, T. Kundu, N. Takeda, Improving accuracy of acoustic source localization in anisotropic plates, Ultrasonics. 54 (2014) 1776–1788.

[8] A. Mostafapour, S. Davoodi, M. Ghareaghaji, Acoustic emission source location in plates using wavelet analysis and cross time frequency spectrum, Ultrasonics. 54 (2014) 2055–2062.

[9]     T. Kundu, X. Yang, H. Nakatani, N. Takeda, A two-step hybrid technique for accurately localizing acoustic source in anisotropic structures without knowing their material properties, Ultrasonics. 56 (2015) 271–278.

[10]    M. Kabir, H. Saboonchi, D. Ozevin, Accurate Source Localization Using Highly Narrowband and Densely Populated MEMS Acoustic Emission Sensors, in: Proc. IWSHM 2015, Destech Publications, Stanford, 2015.

[11]    F. Zahedi, J. Yao, H. Huang, A passive wireless ultrasound pitch–catch system, Smart Mater. Struct. 24 (2015) 85030.

[12]    S.A. Hoseini Sabzevari, M. Moavenian, Locating the acoustic source in thin glass plate using low sampling rate data, Ultrasonics. 70 (2016) 1–11.

[13]    W.H. Park, P. Packo, T. Kundu, Acoustic source localization in an anisotropic plate without knowing its material properties: a new approach, in: T. Kundu (Ed.), Proc. SPIE, Las Vegas, 2016: p. 98050J.

[14]    K. Grabowski, M. Gawronski, I. Baran, W. Spychalski, W.J. Staszewski, T. Uhl, et al., Time–distance domain transformation for Acoustic Emission source localization in thin metallic plates, Ultrasonics. 68 (2016) 142–149.

[15]    H. Achdjian, E. Moulin, F. Benmeddour, J. Assaad, L. Chehami, Source Localisation in a Reverberant Plate Using Average Coda Properties and Early Signal Strength, Acta Acust.

United with Acust. 100 (2014) 834–841.

[16] R. Ernst, F. Zwimpfer, J. Dual, One sensor acoustic emission localization in plates, Ultrasonics. 64 (2016) 139–150.

[17] F. Ciampa, M. Meo, Acoustic emission localization in complex dissipative anisotropic structures using a one-channel reciprocal time reversal method, J. Acoust. Soc. Am. 130 (2011) 168.

[18] M. Surgeon, M. Wevers, One sensor linear location of acoustic emission events using plate wave theories, Mater. Sci. Eng. A. 265 (1999) 254–261.

[19] N. Toyama, J.-H. Koo, R. Oishi, M. Enoki, T. Kishi, Two-dimensional AE source location with two sensors in thin CFRP plates, J. Mater. Sci. Lett. 20 (2001) 1823–1825.

[20] J. Jiao, C. He, B. Wu, R. Fei, X. Wang, Application of wavelet transform on modal acoustic emission source location in thin plates with one sensor, Int. J. Press. Vessel. Pip. 81 (2004) 427–431.

[21] J. Jiao, B. Wu, C. He, Acoustic emission source location methods using mode and frequency analysis, Struct. Control Heal. Monit. 15 (2008) 642–651.

[22] B. Park, H. Sohn, S.E. Olson, M.P. DeSimio, K.S. Brown, M.M. Derriso, Impact localization in complex structures using laser-based time reversal, Struct. Heal. Monit. 11 (2012) 577–588.

[23]   S. Mallat, A wavelet tour of signal processing, Third, Elsevier Inc., 1998.

[24]   N.C. Tse, L. Lai, Wavelet-Based Algorithm for Signal Analysis, EURASIP J. Adv. Signal Process. 2007 (2007) 38916.

[25]   J.B. Harley, J.M.F. Moura, Sparse recovery of the multimodal and dispersive characteristics of Lamb waves, J. Acoust. Soc. Am. 133 (2013) 2732–2745.

[26]   R.M. Levine, J.E. Michaels, Model-based imaging of damage with Lamb waves via sparse reconstruction, J. Acoust. Soc. Am. 133 (2013) 1525.

[27]   A. Ebrahimkhanlou, B. Dubuc, S. Salamone, Damage localization in metallic plate structures using edge-reflected lamb waves, Smart Mater. Struct. 25 (2016) 85035.

[28]   A. Muller, B. Robertson-Welsh, P. Gaydecki, M. Gresil, C. Soutis, Structural Health Monitoring Using Lamb Wave Reflections and Total Focusing Method for Image Reconstruction, Appl. Compos. Mater. (2016) 1–21.

[29]   P.J. Torvik, Reflection of wave trains in semi-infinite plates, J. Acoust. Soc. Am. 41 (1967) 346–353.

[30]   A. Gunawan, S. Hirose, Reflection of Obliquely Incident Guided Waves by an Edge of a Plate, Mater. Trans. 48 (2007) 1236–1243.

[31]   P.D. Wilcox, Lamb wave inspection of large structures using permanently attached transducers, Imperial College of Science, Technology and Medicine, 1998.

[32]   M. Abramowitz, I.A. Stegun, Handbook of mathematical functions: with formulas, graphs, and mathematical tables, Courier Corporation, 1964.

[33]   A. Ebrahimkhanlou, B. Dubuc, S. Salamone, A guided ultrasonic imaging approach in isotropic plate structures using edge reflections, in: J.P. Lynch (Ed.), Proc. SPIE, Sensors Smart Struct. Technol. Civil, Mech. Aerosp. Syst., SPIE, Las Vegas, 2016: p. 98033I.

[34]   N. Quaegebeur, P. Masson, D. Langlois-Demers, P. Micheau, Dispersion-based imaging for structural health monitoring using sparse and compact arrays, Smart Mater. Struct. 20 (2011) 25005.

[35]   B. Park, H. Sohn, S.E. Olson, M.P. DeSimio, K.S. Brown, M.M. Derriso, Impact localization in complex structures using laser-based time reversal, Struct. Heal. Monit. 11 (2012) 577–588.

[36]   J.H. Kurz, C.U. Grosse, H.-W. Reinhardt, Strategies for reliable automatic onset time picking of acoustic emissions and of ultrasound signals in concrete, Ultrasonics. 43 (2005) 538–546.